\documentclass[a4paper]{article}
\usepackage{url, graphicx, color, varioref, amsfonts, longtable, float}
\newcommand\ASTART{\bigskip\noindent\begin{minipage}[c]{0.5\linewidth}}

\newcommand\AENDSKIP{\end{minipage}\bigskip}
\newcommand\AEND{\end{minipage}}

\newcommand{\qed}{\nobreak \ifvmode \relax \else
      \ifdim\lastskip<1.5em \hskip-\lastskip
      \hskip1.5em plus0em minus0.5em \fi \nobreak
      \vrule height0.75em width0.5em depth0.25em\fi}

\title{A Bounded-error Quantum Polynomial Time Algorithm for Two Graph Bisection Problems}

\author{Ahmed Younes\footnote {ayounes2@yahoo.com or ayounes@alexu.edu.eg}\\
Department of Mathematics and Computer Science, \\Faculty of Science, Alexandria University, \\Alexandria, Egypt 
\and Honorary Research Fellow, School of Computer Science,\\ University of Birmingham, Birmingham, B15 2TT, United Kingdom}

\begin{document}
\maketitle

\begin{abstract}
The aim of the paper is to propose a bounded-error quantum polynomial time (BQP) algorithm for the max-bisection and 
the min-bisection problems. The max-bisection and the min-bisection problems are fundamental NP-hard problems. 
Given a graph with even number of vertices, the aim of the max-bisection problem is to divide 
the vertices into two subsets of the same size to maximize the number of edges between the two subsets, 
while the aim of the min-bisection problem is to minimize the number of edges between the two subsets. 
The proposed algorithm runs in $O(m^2)$ for a graph with $m$ edges and in the worst case runs in $O(n^4)$ for a dense 
graph with $n$ vertices. The proposed algorithm targets a general graph by representing both problems as Boolean 
constraint satisfaction problems where the set of satisfied constraints are simultaneously maximized/minimized 
using a novel iterative partial negation and partial measurement technique. 
The algorithm is shown to achieve an arbitrary high probability of success of 
$1-\epsilon$ for small $\epsilon>0$ using a polynomial space resources.

%
%PACS{03.67.Ac,03.67.Lx,03.65.Yz} % end of PACS codes

\end{abstract}

\section{Introduction}

Given an undirected graph $G = (V ,E)$ with a set $V$ of even number of vertices and a set $E$ of unweighted edges. 
Two graph bisection problems will be considered in the paper, the max-bisection problem and the min-bisection problem.
The goal of the max-bisection problem is to divide $V$ into two subsets $A$ and $B$ of the same size 
so as to maximize the number of edges between $A$ and $B$, while the goal of the min-bisection problem is to 
minimize the number of edges between $A$ and $B$. In theory, both bisection problems are NP-hard for general graphs \cite{Ref26,NPhard2007}.

These classical combinatorial optimization problems are special cases of graph partitioning \cite{Ref25}. 
The graph partitioning has many applications, for example, divide-and-conquer algorithms \cite{Ref50}, 
compiler optimization \cite{Ref39}, VLSI circuit layout \cite{Ref8}, load balancing \cite{Ref33}, image processing \cite{Ref62}, 
computer vision \cite{Ref46}, distributed computing \cite{Ref51}, and route planning \cite{Ref19}. In practice, 
there are many general-purpose heuristics for graph partitioning, e.g. \cite{Ref31,Ref58,Ref59} 
that handle particular graph classes, such as \cite{Ref53,Ref16,Ref59}. There are also many practical exact algorithms 
for graph bisection that use the branch-and-bound approach \cite{Ref47,Delling2014}. 
These approaches make expensive usage of time and space to obtain lower bounds \cite{Ref1,Ref3,Ref30,Delling2014}.

On conventional computers, approximation algorithms have gained much attention to tackle the max-bisection and the min-bisection problems.
The max-bisection problem, for example, has an approximation ratio of 0.7028 due to \cite{Feige2006} which is known to be 
the best approximation ratio for a long time by introducing the RPR$^2$ rounding technique into semidefinite programming (SDP) 
relaxation. In \cite{Guruswami2011}, a poly-time algorithm is proposed that, given a graph admitting a bisection cutting a fraction 
$1-\varepsilon$ of edges, finds a bisection cutting an $(1-g(\varepsilon))$ fraction of edges where 
$g(\varepsilon) \to 0$ as $\varepsilon \to 0$. A 0.85-approximation algorithm for the max-bisection is obtained in 
\cite{Raghavendra2012}. In \cite{ZiXu2014}, the SDP relaxation and the RPR$^2$ technique of \cite{Feige2006} 
have been used to obtain a performance curve as a function of the ratio of the optimal SDP value over the total
weight through finer analysis under the assumption of convexity of the RPR$^2$ function. For the min-bisection problem, 
the best known approximation ratio is $O(log\,n)$ \cite{Ref55} 
with some limited graph classes have known polynomial-time solutions such as grids without holes \cite{Ref22}
and graphs with bounded tree width \cite{Ref37}.

The aim of the paper is to propose an algorithm that represents the two bisection problems as Boolean constraint 
satisfaction problems where the set of edges are represented as set of constraints. 
The algorithm prepares a superposition of all possible graph bisections using an amplitude 
amplification technique then evaluates the set of constraints for all possible bisections simultaneously and then amplifies 
the amplitudes of the best bisections that achieve the maximum/minimum satisfaction to the set of constraints 
using a novel amplitude amplification technique that applies an iterative partial negation and partial measurement. 
The proposed algorithm targets a general graph where it runs in $O(m^2)$ for a graph with $m$ edges and in the worst case 
runs in $O(n^4)$ for a dense graph with number edges close to $m = {\textstyle{{n(n - 1)} \over 2}}$ with $n$ vertices 
to achieve an arbitrary high probability of success of $1-\epsilon$ for small $\epsilon>0$ using a polynomial space resources.

The paper is organized as follows; Section 2 shows the data structure used to represent a graph bisection problem 
as a Boolean constraint satisfaction 
problem. Section 3 presents the proposed algorithm with analysis on time and space requirements. 
Section 4 concludes the paper. 

%======================================================================================

\section{Data Structures and Graph Representation}

Several optimization problems, such as the max-bisection and the min-bisection problems, can be formulated as Boolean constraint satisfaction
problems \cite{CombOptBook,BAZ05} where a feasible solution is a solution with as many variables set to 0
as variables set to 1, i.e. balanced assignment, as follows: 
for a graph $G$ with $n$ vertices and $m$ edges, consider $n$ Boolean variables $v_0, \ldots , v_{n-1}$
and $m$ constraints by associating with each edge $(a, b) \in E$ the constraint $c_l=v_a\oplus v_b$, with $l=0,1,\ldots,m-1$, 
then the max-bisection is the problem that consists of finding a balanced assignment 
to maximize the number of constraints equal to logic-1 from the $m$ constraints, 
while the min-bisection is the problem that consists of finding a balanced assignment to 
maximize the number of constraints equal to logic-0 from the $m$ constraints, such that 
if a Boolean variable is set to 0 then the associated vertex belongs to the first partition and 
if a Boolean variable is set to 1 then the associated vertex belongs to the second partition.

\begin{center}
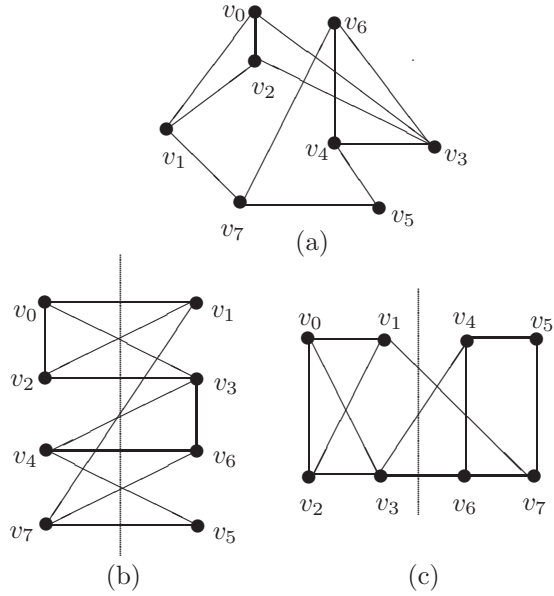
\begin{figure*}[htbp]
\begin{center}

\setlength{\unitlength}{3947sp}%
\begingroup\makeatletter\ifx\SetFigFont\undefined
% extract first six characters in \fmtname
\def\x#1#2#3#4#5#6#7\relax{\def\x{#1#2#3#4#5#6}}%
\expandafter\x\fmtname xxxxxx\relax \def\y{splain}%
\ifx\x\y   % LaTeX or SliTeX?
\gdef\SetFigFont#1#2#3{%
  \ifnum #1<17\tiny\else \ifnum #1<20\small\else
  \ifnum #1<24\normalsize\else \ifnum #1<29\large\else
  \ifnum #1<34\Large\else \ifnum #1<41\LARGE\else
     \huge\fi\fi\fi\fi\fi\fi
  \csname #3\endcsname}%
\else
\gdef\SetFigFont#1#2#3{\begingroup
  \count@#1\relax \ifnum 25<\count@\count@25\fi
  \def\x{\endgroup\@setsize\SetFigFont{#2pt}}%
  \expandafter\x
    \csname \romannumeral\the\count@ pt\expandafter\endcsname
    \csname @\romannumeral\the\count@ pt\endcsname
  \csname #3\endcsname}%
\fi
\fi\endgroup
\begin{picture}(3346,3654)(1090,-3361)
\thinlines
\put(2616,240){\circle*{90}}
\put(2064,-489){\circle*{90}}
\put(3733,-605){\circle*{90}}
\put(2523,-951){\circle*{90}}
\put(3387,-984){\circle*{90}}
\put(2614,-66){\circle*{90}}
\put(3109,173){\circle*{90}}
\put(3109,-579){\circle*{90}}
\put(2950,-2668){\circle*{90}}
\put(2945,-1800){\circle*{90}}
\put(3421,-1814){\circle*{90}}
\put(4364,-1807){\circle*{90}}
\put(3396,-2662){\circle*{90}}
\put(3916,-2665){\circle*{90}}
\put(3935,-1818){\circle*{90}}
\put(4351,-2667){\circle*{90}}
\put(1308,-1573){\circle*{90}}
\put(1312,-2050){\circle*{90}}
\put(1318,-2500){\circle*{90}}
\put(1319,-2966){\circle*{90}}
\put(2261,-2976){\circle*{90}}
\put(2254,-2508){\circle*{90}}
\put(2255,-2056){\circle*{90}}
\put(2253,-1579){\circle*{90}}
\put(2608,243){\line(-3,-4){564}}
\put(2611,243){\line( 4,-3){1128}}
\put(2058,-481){\line( 1,-1){470}}
\put(3599,-63){\makebox(1.6667,11.6667){\SetFigFont{5}{6}{rm}.}}
\put(2618,233){\line( 0,-1){282}}
\put(3724,-596){\line(-2, 1){1128}}
\put(3112,158){\line( 0,-1){752}}
\put(3375,-975){\line(-2, 3){282}}
\put(2535,-960){\line( 1, 2){564}}
\put(3698,-580){\line(-3, 4){564}}
\put(3698,-586){\line(-1, 0){564}}
\put(3378,-968){\line(-1, 0){846}}
\put(2061,-494){\line( 4, 3){564}}
\put(2955,-1807){\line( 1, 0){470}}
\put(2950,-1812){\line( 0,-1){846}}
\put(2960,-1807){\line( 1,-2){423}}
\put(3439,-1794){\line( 1,-1){893}}
\put(3395,-1820){\line(-1,-2){423}}
\put(2965,-2650){\line( 1, 0){423}}
\put(3384,-2650){\line( 2, 3){564}}
\put(3384,-2660){\line( 1, 0){517}}
\put(3948,-1797){\line( 1, 0){423}}
\put(3924,-2665){\line( 0, 1){846}}
\put(4367,-1807){\line( 0,-1){846}}
\put(3905,-2660){\line( 1, 0){470}}
\multiput(3630,-1498)(0.00000,-11.94915){119}{\makebox(1.6667,11.6667){\SetFigFont{5}{6}{rm}.}}
\put(1307,-1575){\line( 1, 0){940}}
\put(1312,-1580){\line( 0,-1){470}}
\put(1322,-1585){\line( 2,-1){940}}
\put(2237,-1570){\line(-2,-1){940}}
\put(2242,-1580){\line(-2,-3){940}}
\put(1321,-2052){\line( 1, 0){940}}
\put(2259,-2053){\line(-2,-1){940}}
\put(2252,-2052){\line( 0,-1){470}}
\put(1322,-2505){\line( 1, 0){940}}
\put(1331,-2510){\line( 2,-1){940}}
\put(1317,-2972){\line( 1, 0){940}}
\put(2245,-2515){\line(-2,-1){940}}
\multiput(1779,-1286)(0.00000,-11.97452){158}{\makebox(1.6667,11.6667){\SetFigFont{5}{6}{rm}.}}
\put(2039,-708){$v_1$}
\put(3790,-692){$v_3$}
\put(2401,-1162){$v_7$}
\put(3461,-1073){$v_5$}
\put(2595,-270){$v_2$}
\put(3168,138){$v_6$}
\put(2927,-662){$v_4$}
\put(2410,210){$v_0$}
\put(2883,-1720){$v_0$}
\put(3379,-1720){$v_1$}
\put(3842,-1715){$v_4$}
\put(4324,-1735){$v_5$}
\put(2893,-2886){$v_2$}
\put(3365,-2877){$v_3$}
\put(3842,-2881){$v_6$}
\put(4276,-2877){$v_7$}
\put(1112,-1666){$v_0$}
\put(1095,-2100){$v_2$}
\put(1110,-2582){$v_4$}
\put(1090,-3050){$v_7$}
\put(2353,-3059){$v_5$}
\put(2353,-2582){$v_6$}
\put(2358,-2129){$v_3$}
\put(2343,-1671){$v_1$}
\put(1688,-3339){(b)}
\put(2864,-1248){(a)}
\put(3562,-3344){(c)}
\end{picture}%

\end{center}
\caption{(a) A random graph with 8 vertices and 12 edges, (b) A max-bisection instance for the graph in (a) with 10 edges connecting 
the two subsets, and (c) (b) A min-bisection instance for the graph in (a) with 3 edges connecting 
the two subsets.}
\label{graphex}
\end{figure*}
\end{center}

For example, consider the graph $G$ shown in Figure \ref{graphex}(a). Let $G = (V ,E)$, where,
\begin{equation}
\begin{array}{l}
V=\{0,1,2,3,4,5,6,7\},\\ 
E=\{(0,1),(0,2),(0,3),\\
\,\,\,\,\,\,\,\,\,\,\,\,\,\,\,\,(1,2),(1,7),(2,3),\\
\,\,\,\,\,\,\,\,\,\,\,\,\,\,\,\,(3,4),(3,6),(4,5),\\
\,\,\,\,\,\,\,\,\,\,\,\,\,\,\,\,(4,6),(5,7),(6,7)\}.\\ 
\end{array}
\end{equation}

Assume that each vertex $a \in V$ is associated 
with a Boolean variable $v_a$, then the set of vertices $V$ can be represented as a vector $X$ of Boolean variables as follows, 
\begin{equation}
X=(v_0,v_1,v_2,v_3,v_4,v_5,v_6,v_7),
\end{equation}
\noindent  
and if each edge $(a, b) \in E$ is associated with a constraint $c_l=v_a\oplus v_b$ 
then the set of edges $E$ can be represented as a vector $Z$ of constraints as follows,

\begin{equation}
Z=(c_0,c_1,c_2,c_3,c_4,c_5,c_6,c_7,c_8,c_9,c_{10},c_{11}), 
\end{equation}
\noindent
such that,
\begin{equation}
\begin{array}{l}
c_0= (v_0\oplus v_1), c_1= (v_0\oplus v_2), c_2= (v_0\oplus v_3),\\ 
c_3= (v_1\oplus v_2), c_4= (v_1\oplus v_7), c_5= (v_2\oplus v_3), \\
c_6= (v_3\oplus v_4), c_7= (v_3\oplus v_6), c_8= (v_4\oplus v_5), \\
c_9= (v_4\oplus v_6), c_{10}= (v_5 \oplus v_7), c_{11}= (v_6\oplus v_7).\\ 
\end{array}
\end{equation}

In general, a bisection $G_P$ for the graph $G$ can be represented as $G_p = (x ,z(x))$ such that 
each vector $x \in \{0,1\}^n $ of variable assignments is associated with a vector $z(x)\in \{0,1\}^m$ of constraints 
evaluated as functions of the variable assignment $x$. In the max-bisection and the min-bisection problems, 
the vector $x$ of variable assignments are restricted to be balanced so there are 
$M = \left( {\begin{array}{*{20}c}
   n  \\
   {{\textstyle{n \over 2}}}  \\
\end{array}} \right)$ possible variable assignments among the $N=2^n$ possible variable assignments, 
and the solution of the max-bisection problem is to find 
the variable assignment that is associated with a vector of constraints that contains the maximum number of 1's, and 
the solution for the min-bisection problem is to find 
the variable assignment that is associated with a vector of constraints that contains the maximum number of 0's. 
For example, for the graph $G$ shown in Figure \ref{graphex}(a), a max-bisection for $G$ is 
$((0,1,0,1,0,1,1,0),(1,0,1,1,1,1,1,0,1,1,1,1))$ with 10 edges connecting the two partitions as shown in Figure \ref{graphex}(b), 
and a min-bisection for $G$ is 
$((0,0,0,0,1,1,1,1),(0,0,0,0,1,0,1,1,0,0,0,0))$ with 3 edges connecting the two partitions as shown in Figure \ref{graphex}(c). 
It is important to notice that a variable assignment $x= (0,1,0,1,0,1,1,0)$ is equivalent to 
$\overline x= (1,0,1,0,1,0,0,1)$, where $\overline x$ is the bit-wise negation of $x$.

\section{The Algorithm}

Given a Graph $G$ with $n$ vertices and $m$ edges. The proposed algorithm is divided into three stages, the first stage prepares 
a superposition of all balanced assignments for the $n$ variables. The second stage evaluates the $m$ constraints 
associated with the $m$ edges for every balanced assignment and stores the values of constraints 
in constraint vectors entangled with the corresponding balanced assignments in the superposition.
The third stage amplifies the constraint vector with maximum (minimum) number of satisfied constraints using 
a partial negation and iterative measurement technique.

\begin{center}
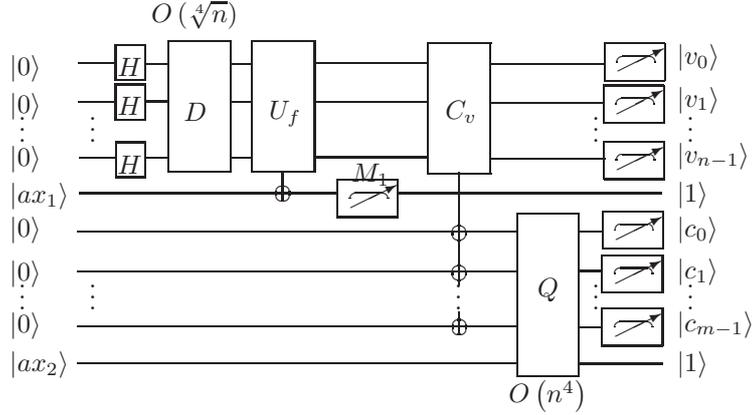
\begin{figure*}[htbp]
\begin{center}

\setlength{\unitlength}{3947sp}%
\begingroup\makeatletter\ifx\SetFigFont\undefined
% extract first six characters in \fmtname
\def\x#1#2#3#4#5#6#7\relax{\def\x{#1#2#3#4#5#6}}%
\expandafter\x\fmtname xxxxxx\relax \def\y{splain}%
\ifx\x\y   % LaTeX or SliTeX?
\gdef\SetFigFont#1#2#3{%
  \ifnum #1<17\tiny\else \ifnum #1<20\small\else
  \ifnum #1<24\normalsize\else \ifnum #1<29\large\else
  \ifnum #1<34\Large\else \ifnum #1<41\LARGE\else
     \huge\fi\fi\fi\fi\fi\fi
  \csname #3\endcsname}%
\else
\gdef\SetFigFont#1#2#3{\begingroup
  \count@#1\relax \ifnum 25<\count@\count@25\fi
  \def\x{\endgroup\@setsize\SetFigFont{#2pt}}%
  \expandafter\x
    \csname \romannumeral\the\count@ pt\expandafter\endcsname
    \csname @\romannumeral\the\count@ pt\endcsname
  \csname #3\endcsname}%
\fi
\fi\endgroup
\begin{picture}(4341,2349)(579,-1764)

\thinlines
\put(2750,-477){\oval(236, 72)[tr]}
\put(2750,-477){\oval(236, 72)[tl]}
\put(2554,-360){\line( 0,-1){235}}
\put(2554,-595){\line( 1, 0){376}}
\put(2930,-595){\line( 0, 1){235}}
\put(2930,-360){\line(-1, 0){376}}
\put(2639,-535){\vector( 2, 1){282}}
\put(4407,-239){\oval(236, 72)[tr]}
\put(4407,-239){\oval(236, 72)[tl]}
\put(4211,-122){\line( 0,-1){235}}
\put(4211,-357){\line( 1, 0){376}}
\put(4587,-357){\line( 0, 1){235}}
\put(4587,-122){\line(-1, 0){376}}
\put(4296,-297){\vector( 2, 1){282}}
\put(4404,114){\oval(236, 72)[tr]}
\put(4404,114){\oval(236, 72)[tl]}
\put(4208,231){\line( 0,-1){235}}
\put(4208, -4){\line( 1, 0){376}}
\put(4584, -4){\line( 0, 1){235}}
\put(4584,231){\line(-1, 0){376}}
\put(4293, 56){\vector( 2, 1){282}}
\put(4410,377){\oval(236, 72)[tr]}
\put(4410,377){\oval(236, 72)[tl]}
\put(4214,494){\line( 0,-1){235}}
\put(4214,259){\line( 1, 0){376}}
\put(4590,259){\line( 0, 1){235}}
\put(4590,494){\line(-1, 0){376}}
\put(4299,319){\vector( 2, 1){282}}
\put(4399,-676){\oval(236, 72)[tr]}
\put(4399,-676){\oval(236, 72)[tl]}
\put(4203,-559){\line( 0,-1){235}}
\put(4203,-794){\line( 1, 0){376}}
\put(4579,-794){\line( 0, 1){235}}
\put(4579,-559){\line(-1, 0){376}}
\put(4288,-734){\vector( 2, 1){282}}
\put(4395,-950){\oval(236, 72)[tr]}
\put(4395,-950){\oval(236, 72)[tl]}
\put(4199,-833){\line( 0,-1){235}}
\put(4199,-1068){\line( 1, 0){376}}
\put(4575,-1068){\line( 0, 1){235}}
\put(4575,-833){\line(-1, 0){376}}
\put(4284,-1008){\vector( 2, 1){282}}
\put(4394,-1283){\oval(236, 72)[tr]}
\put(4394,-1283){\oval(236, 72)[tl]}
\put(4198,-1166){\line( 0,-1){235}}
\put(4198,-1401){\line( 1, 0){376}}
\put(4574,-1401){\line( 0, 1){235}}
\put(4574,-1166){\line(-1, 0){376}}
\put(4283,-1341){\vector( 2, 1){282}}
\put(2214,-445){\oval(84,92)}
\put(2026,-306){\framebox(383,811){}}
\put(2212,-306){\line( 0,-1){188}}

\put(3310,-693){\oval(84,92)}
\put(3311,-943){\oval(84,92)}
\put(3312,-1282){\oval(84,92)}
\put(946,129){\line( 1, 0){229}}
\put(949,-223){\line( 1, 0){234}}
\put(1180, 19){\framebox(176,215){}}
\put(1180,267){\framebox(176,215){}}
\put(1181,-343){\framebox(176,215){}}

\put(1359,366){\line( 1, 0){141}}
\put(1365,130){\line( 1, 0){139}}
\put(1359,-223){\line( 1, 0){141}}
\put(945,369){\line( 1, 0){229}}
\put(1506,-307){\framebox(383,811){}}
\put(2406,128){\line( 1, 0){705}}
\put(4055,-1282){\line( 1, 0){141}}
\put(3505,122){\line( 1, 0){705}}
\put(3511,-228){\line( 1, 0){705}}
\put(3511,367){\line( 1, 0){705}}
\put(4055,-1512){\line( 1, 0){517}}
\put(3310,-1021){\line( 0, 1){705}}
\put(3310,-1326){\line( 0, 1){141}}
\put(2408,371){\line( 1, 0){705}}
\put(953,-445){\line( 1, 0){1598}}
\put(3119,-319){\framebox(383,811){}}
\put(2411,-220){\line( 1, 0){705}}
\put(942,-1509){\line( 1, 0){2726}}
\put(941,-1276){\line( 1, 0){2726}}
\put(941,-937){\line( 1, 0){2726}}
\put(941,-686){\line( 1, 0){2726}}
\put(4062,-688){\line( 1, 0){141}}
\put(4203,-688){\line(-1, 0){141}}
\put(4058,-931){\line( 1, 0){141}}
\put(4199,-931){\line(-1, 0){141}}
\put(1889,367){\line( 1, 0){135}}
\put(1889,128){\line( 1, 0){135}}
\put(1892,-224){\line( 1, 0){133}}
\put(2923,-445){\line( 1, 0){1645}}
\put(3674,-1589){\framebox(378,1014){}}

\put(1190,-322){$H$}
\put(1190, 46){$H$}
\put(1190,288){$H$}

\put(3226, 19){$C_v$}
\put(2132, 19){$U_f$}
\put(1602,  10){$D$}

\put(3800,-1091){$Q$}

\put(4665,363){$\left| {v_0} \right\rangle$ }
\put(4665,100){$\left| {v_1} \right\rangle $}
\put(4665,-250){$\left| {v_{n-1}} \right\rangle$ }
\put(4665,-1558){$\left| {1} \right\rangle $}
\put(4665,-499){$\left| {1} \right\rangle $}
\put(4665,-714){$\left| {c_0} \right\rangle $}
\put(4665,-977){$\left| {c_1} \right\rangle $}
\put(4665,-1307){$\left| {c_{m-1}} \right\rangle $}

\put(520,297){$\left| {0} \right\rangle $}
\put(520, 58){$\left| {0} \right\rangle $}
\put(520,-274){$\left| {0} \right\rangle $}
\put(520,-505){$\left| {ax_1} \right\rangle $}
\put(520,-714){$\left| {0} \right\rangle $}
\put(520,-989){$\left| {0} \right\rangle $}
\put(520,-1319){$\left| {0} \right\rangle $}
\put(520,-1558){$\left| {ax_2} \right\rangle $}

\put(1400,627){$O\left( {\sqrt[4]{n}} \right)$}
%\put(4100,627){$O\left( {n^2} \right)$}
\put(3620,-1755){$O\left( {{n^4}} \right)$}

\put(520,-1170){ $\vdots$ }
\put(520,-120){ $\vdots$ }
\put(956,-1170){ $\vdots$ }
\put(956,-120){ $\vdots$ }

\put(4081,-1170){ $\vdots$ }
\put(4081,-120){ $\vdots$ }
\put(4670,-120){ $\vdots$ }
\put(4670,-1170){ $\vdots$ }

\put(3230,-1170){ $\vdots$ }

\put(2643,-374){$M_1$}

\end{picture}%

\end{center}
\caption{A quantum circuit for the proposed algorithm.}
\label{alg}
\end{figure*}
\end{center}

\subsection{Balanced Assignments Preparation}

To prepare a superposition of all balanced assignments of $n$ qubits, the proposed algorithm can use any 
amplitude amplification technique, e.g. \cite{Grover1997,Younes2007,Younes2013}. An extra step should be added after 
the amplitude amplification to create an entanglement between the matched items and an auxiliary qubit 
$\left| {ax_1 } \right\rangle$, so that the correctness of the items in the 
superposition can be verified by applying measurement on $\left| {ax_1 } \right\rangle$ without having to examine 
the superposition itself. So, if $\left| {ax_1 } \right\rangle = \left| {1 } \right\rangle$ at the end of 
this stage, then the superposition contains the correct items, i.e. the balanced assignments, 
otherwise, repeat the preparation stage until $\left| {ax_1 } \right\rangle = \left| {1 } \right\rangle$. 
This is useful for not having to proceed to the next stages until the preparation stage succeeds.

The preparation stage to have a superposition of all balanced assignments of $n$ qubits will use the 
amplitude amplification technique shown in \cite{Younes2013} since it achieves the highest known probability of success 
using fixed operators and it can be summarized as follows, prepare a superposition of $2^n$ states by initializing 
$n$ qubits to state $\left| {0} \right\rangle$ and apply $H^{\otimes n}$ on the $n$ qubits
\begin{equation}
\begin{array}{l}
 \left| {\Psi _0 } \right\rangle  = \left( {H^{ \otimes n}} \right) \left| 0 \right\rangle ^{ \otimes n}  \\ 
 \,\,\,\,\,\,\,\,\,\,\,\,\, = \frac{1}{{\sqrt N }}\sum\limits_{j = 0}^{N - 1} {\left| j \right\rangle },  \\ 
 \end{array}
\end{equation}

\noindent
where $H$ is the Hadamard gate, and $N=2^n$. 
Assume that the system $\left| {\Psi _0 } \right\rangle$ is re-written as follows,

\begin{equation}
\label{ENheqn39}
\begin{array}{l}
\left|\Psi_0\right\rangle =  \frac{1}{{\sqrt N }} \sum\limits_{\scriptstyle j = 0, \hfill \atop 
  \scriptstyle j \in X_T  \hfill}^{N - 1} {\left| j \right\rangle} 
  + \frac{1}{{\sqrt N }} \sum\limits_{\scriptstyle j = 0, \hfill \atop 
  \scriptstyle j \in X_F  \hfill}^{N - 1} {\left| j \right\rangle},\\
\end{array} 
\end{equation}

\noindent
where $X_T$ is the set of all balanced assignments of $n$ bits and $X_F$ is the set of all 
unbalanced assignments. Let $M=\left( {\begin{array}{*{20}c}
   n  \\
   {{\textstyle{n \over 2}}}  \\
\end{array}} \right)$ be the number of balanced assignments among the $2^n$ possible assignments, 
$\sin (\theta ) = \sqrt {{M \mathord{\left/ {\vphantom {M N}} \right.\kern-\nulldelimiterspace} N}}$ 
and $0 < \theta  \le \pi /2$, then the system can be re-written as follows,

\begin{equation}
\left|\Psi_0\right\rangle  = \sin (\theta )\left| {\psi _1 } \right\rangle  + \cos (\theta )\left| {\psi _0 } \right\rangle,
\end{equation}

\noindent
where $\left| {\psi _1 } \right\rangle= \left| {\tau  } \right\rangle$ represents the balanced assignments subspace and 
$\left| {\psi _0 } \right\rangle$ represents the unbalanced assignments subspace.

Let $D=WR_0 \left( \phi  \right)W^\dag  R_\tau  \left( \phi  \right)$, 
$R_0 \left( \phi  \right) = I - (1 - e^{i\phi } )\left| 0 \right\rangle \left\langle 0 \right|$, 
$R_\tau  \left( \phi  \right) = I - (1 - e^{i\phi } )\left| \tau  \right\rangle \left\langle \tau  \right|$, 
where $W=H^{\otimes n}$ is the Walsh-Hadamard transform \cite{hoyer00}. 
Iterate the operator $D$ on $\left|\Psi_0\right\rangle$ for $q$ times to get,

\begin{equation}
\left| {\Psi _1 } \right\rangle  = D^{q}\left| {\Psi_0 } \right\rangle  = a_q \left| {\psi _1 } \right\rangle  + b_q \left| {\psi _0 } \right\rangle ,
\end{equation}

\noindent
such that, 
\begin{equation}
\label{aqeqn}
a_q  = \sin (\theta )\left( {e^{iq\phi } U_q \left( y \right) + e^{i(q - 1)\phi } U_{q - 1} \left( y \right)} \right), 
\end{equation}

\begin{equation}
b_q =  \cos (\theta )e^{i(q - 1)\phi } \left( {U_q \left( y \right) + U_{q - 1} \left( y \right)} \right),
\end{equation} 

\noindent  
where $y=cos(\delta)$, $\cos \left( \delta  \right) = 2\sin ^2 (\theta )\sin ^2 ({\textstyle{\phi  \over 2}}) - 1$,
 $0<\theta\le \pi/2$, and $U_q$ is the Chebyshev polynomial of the second kind \cite{ChebPoly} defined as follows, 

\begin{equation}
 U_q \left( y \right) = \frac{{\sin \left( {\left( {q + 1} \right)\delta } \right)}}{{\sin \left( \delta  \right)}}.
\end{equation}

Setting $\phi=6.02193\approx1.9168\pi$, $M=\left( {\begin{array}{*{20}c}
   n  \\
   {{\textstyle{n \over 2}}}  \\
\end{array}} \right)$, $N=2^n$ and, $q = \left\lfloor {{\textstyle{\phi  \over {\sin (\theta)}}}} \right\rfloor$, then
$\left| {a_q } \right|^2  \ge 0.9975$ \cite{Younes2013}. The upper bound for the required number of iterations $q$ to reach the maximum probability of success is,

\begin{equation}
\label{ENheqn64}
q = \left\lfloor {{\textstyle{\phi  \over {\sin (\theta)}}}} \right\rfloor \le 1.9168\pi\sqrt {\frac{N}{M}},
\end{equation}
\noindent
and using Stirling's approximation,

\begin{equation}
n! \approx \sqrt {2\pi n} \left( {\frac{n}{e}} \right)^n,
\end{equation}
\noindent
then, the upper bound for required number of iterations $q$ to prepare the superposition of all balanced assignments is,

\begin{equation}
q \approx 1.9168\sqrt[4]{{\frac{{\pi ^5 }}{{2 }}n}} = O\left( \sqrt[4]{n} \right).
\end{equation}

It is required to preserve the states in $\left| {\psi _1 } \right\rangle$ for further processing in the next stage. 
This can be done by adding an auxiliary qubit $\left| {ax_1 } \right\rangle$ initialized to state 
$\left| {0} \right\rangle$ and have the states of the balanced assignments entangled with 
$\left| {ax_1 } \right\rangle = \left| {1} \right\rangle$, so that, 
the correctness of the items in the 
superposition can be verified by applying measurement on $\left| {ax_1 } \right\rangle$ without having to examine 
the superposition itself. So, if $\left| {ax_1 } \right\rangle = \left| {1 } \right\rangle$, 
then the superposition contains the balanced assignments, 
otherwise, repeat the preparation stage until $\left| {ax_1 } \right\rangle = \left| {1 } \right\rangle$. 
This is useful to be able to proceed to the next stage when the preparation stage succeeds. To prepare the entanglement, let

\begin{equation}
\begin{array}{l}
\left| {\Psi _2 } \right\rangle = \left| {\Psi _1 } \right\rangle \otimes \left| {0 } \right\rangle\\  
\,\,\,\,\,\,\,\,\,\,\,\,\,= a_q \left| {\psi _1 } \right\rangle \otimes \left| {0 } \right\rangle + b_q \left| {\psi _0 } \right\rangle \otimes \left| {0 } \right\rangle,\\
\end{array}
\end{equation}
\noindent 
and apply a quantum Boolean operator $U_f$ on $\left| {\Psi _2 } \right\rangle$, where $U_f$ is defined as follows,

\begin{equation}
U_f \left| {x,0} \right\rangle  = \left\{ {\begin{array}{*{20}c}
   {\left| {x ,0} \right\rangle ,{\rm if }\, \left|x\right\rangle \in \left|\psi _0\right\rangle,}  \\
   {\left| {x ,1} \right\rangle ,{\rm if }\, \left|x\right\rangle \in \left|\psi _1\right\rangle,}  \\
\end{array}} \right.
\end{equation}

\noindent
and $f:\left\{ {0,1} \right\}^n  \to \{ 0,1\}$ is an $n$ inputs single output Boolean function that 
evaluates to True for any $x \in X_T$ and evaluates to False for any $x \in X_F$, then,

\begin{equation}
\begin{array}{l}
\left| {\Psi _3 } \right\rangle = U_f \left| {\Psi _2 } \right\rangle \\  
\,\,\,\,\,\,\,\,\,\,\,\,\,= a_q \left| {\psi _1 } \right\rangle \otimes \left| {1 } \right\rangle + b_q \left| {\psi _0 } \right\rangle \otimes \left| {0 } \right\rangle.\\
\end{array}
\end{equation}

Apply measurement $M_1$ on the auxiliary qubit $\left| {ax_1 } \right\rangle$ as shown in Figure \ref{alg}. 
The probability of finding $\left| {ax_1 } \right\rangle=\left| {1} \right\rangle$ is,

\begin{equation}
Pr{(M_1  = 1)}  = \left| {a_q } \right|^2  \ge 0.9975,
\end{equation}

\noindent
and the system will collapse to, 
 
\begin{equation}
\left|\Psi_3^{(M_1  = 1)}\right\rangle = \left| {\psi _1 } \right\rangle \otimes \left| {1 } \right\rangle. 
\end{equation}

 \subsection{Evaluation of Constraints}
 
There are $M$ states in the superposition $\left|\Psi_3^{(M_1  = 1)}\right\rangle$, each state has an amplitude 
${\textstyle{1 \over {\sqrt M }}}$, then let $\left|\Psi_4\right\rangle$ be the system after 
the balanced assignment preparation stage as follows, 

\begin{equation}
\left|\Psi_4\right\rangle = \alpha \sum\limits_{k = 0}^{M - 1} {\left| x_k \right\rangle},
\end{equation}

\noindent
where $\left|ax_1\right\rangle$ is dropped from the system for simplicity and $\alpha = {\textstyle{1 \over {\sqrt M }}}$. 
For a graph $G$ with $n$ vertices and $m$ edges, every edge $(a, b )$ connecting vertcies $a,b \in V$ 
is associated with a constraint $c_l=v_a\oplus v_b$, where $v_a$ and $v_b$ are the corresponding qubits for vertices $a$ and $b$ 
in $\left|\Psi_4\right\rangle$ respectively such that 
$0 \le l < m$, $0 \le m \le {\textstyle{{n(n - 1)} \over 2}}$, $0 \le a,b \le n-1$ and $a \ne b$, where 
${\textstyle{{n(n - 1)} \over 2}}$ is the maximum number of edges in a graph with $n$ vertices.

To evaluate the $m$ constraints associated with the edges, add $m$ qubits initialized 
to state $\left|0\right\rangle$,   

\begin{equation}
\begin{array}{l}
\left|\Psi_5\right\rangle = \left|\Psi_4\right\rangle \otimes \left|0\right\rangle^{\otimes m}\\
\,\,\,\,\,\,\,\,\,\,\,\,\,=\alpha \sum\limits_{k = 0}^{M - 1} { {\left| x_k \right\rangle \otimes \left|0 \right\rangle^{\otimes m} } }.\\
\end{array}
\end{equation}

For every constraint $c_l=v_a \oplus v_b$, apply two $Cont\_\sigma_X$ gates, $Cont\_\sigma_X(v_a,c_l)$ and $Cont\_\sigma_X(v_b,c_l)$, so that 
$\left|c_l\right\rangle=\left|v_a\oplus v_b\right\rangle$. The collection of all 
$Cont\_\sigma_X$ gates applied to evaluate the $m$ constraints is denoted $C_v$ in Figure \ref{alg}, 
then the system is transformed to,

\begin{equation}
\left| {\Psi _6 } \right\rangle  = \alpha \sum\limits_{k = 0}^{M - 1} 
{\left( {\left| x_k \right\rangle  \otimes \left| {c_0^k c_1^k  \ldots c_{m-1}^k } \right\rangle } \right)}, 
\end{equation}
\noindent
where $\sigma _X$ is the Pauli-X gate which is the quantum equivalent to 
the NOT gate. It can be seen as a rotation of the Bloch Sphere around the X-axis by $\pi$ radians as follows,

\begin{equation}
\sigma _X  = \left[ {\begin{array}{*{20}c}
   0 & 1  \\
   1 & 0  \\
\end{array}} \right],
\end{equation}
and $Cont\_U(v,c)$ gate is a controlled gate with control qubit $\left|v\right\rangle$ and target qubit $\left|c\right\rangle$ 
that applies a single qubit unitary operator $U$ on $\left|c\right\rangle$ only if $\left|v\right\rangle=\left|1\right\rangle$, 
so every qubit $\left|c_l^k\right\rangle$ carries a value of the constraint $c_l$ based on the values of 
$v_a$ and $v_b$ in the balanced assignment $\left| x_k \right\rangle$, i.e. the values of $v_a^k$ and $v_b^k$ respectively. 
Let $\left| z_k \right\rangle=\left|{c_0^k c_1^k \ldots c_{m-1}^k }\right\rangle$, then the system can 
be re-written as follows,

\begin{equation}
\left| {\Psi _6 } \right\rangle  = \alpha \sum\limits_{k = 0}^{M - 1} 
{\left( {\left| x_k \right\rangle  \otimes \left| {z_k} \right\rangle } \right)}, 
\end{equation}
\noindent
where every $\left| x_k \right\rangle$ is entangled with the corresponding $\left| z_k \right\rangle$. 
The aim of the next stage is to find $\left| z_k \right\rangle$ with the maximum number of 
$\left| 1 \right\rangle$'s for the max-bisection problem or to find $\left| z_k \right\rangle$ with 
the minimum number of $\left| 1 \right\rangle$'s for the min-bisection problem.

\begin{center}
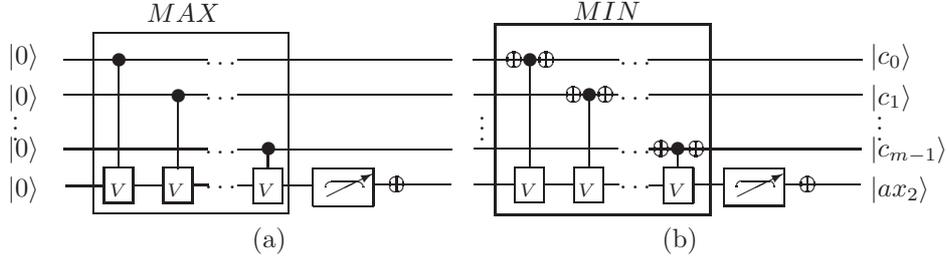
\begin{figure*}[htbp]
\begin{center}

\setlength{\unitlength}{3947sp}%
\begingroup\makeatletter\ifx\SetFigFont\undefined
% extract first six characters in \fmtname
\def\x#1#2#3#4#5#6#7\relax{\def\x{#1#2#3#4#5#6}}%
\expandafter\x\fmtname xxxxxx\relax \def\y{splain}%
\ifx\x\y   % LaTeX or SliTeX?
\gdef\SetFigFont#1#2#3{%
  \ifnum #1<17\tiny\else \ifnum #1<20\small\else
  \ifnum #1<24\normalsize\else \ifnum #1<29\large\else
  \ifnum #1<34\Large\else \ifnum #1<41\LARGE\else
     \huge\fi\fi\fi\fi\fi\fi
  \csname #3\endcsname}%
\else
\gdef\SetFigFont#1#2#3{\begingroup
  \count@#1\relax \ifnum 25<\count@\count@25\fi
  \def\x{\endgroup\@setsize\SetFigFont{#2pt}}%
  \expandafter\x
    \csname \romannumeral\the\count@ pt\expandafter\endcsname
    \csname @\romannumeral\the\count@ pt\endcsname
  \csname #3\endcsname}%
\fi
\fi\endgroup
\begin{picture}(5618,2421)(585,-1902)
\thinlines

\put(1120,-1674){\framebox(1208,1124){}}
\put(3616,-1684){\framebox(1331,1187){}}

\put(5945,-1573){$\left|ax_2\right\rangle$}
\put(5945,-1323){$\left|c_{m-1}\right\rangle$}
\put(5945,-994){{$\left|c_1\right\rangle$}}
\put(5945,-740){{$\left|c_0\right\rangle$}}
\put(585,-1573){$\left|0\right\rangle$}
\put(585,-1323){$\left|0\right\rangle$}
\put(585,-994){$\left|0\right\rangle$}
\put(585,-740){$\left|0\right\rangle$}
\put(1220,-1545){$\mathop {}\nolimits_{V }$}
\put(1578,-1545){$\mathop {}\nolimits_{V }$}
\put(2139,-1545){$\mathop {}\nolimits_{V }$}
\put(3768,-1545){$\mathop {}\nolimits_{V }$}
\put(4131,-1545){$\mathop {}\nolimits_{V }$}
\put(4692,-1545){$\mathop {}\nolimits_{V }$}
\put(610,-1220){$\vdots$}
\put(3509,-1220){$\vdots$}
\put(5980,-1220){$\vdots$}
%\put(1593, 28){\makebox(1.6667,11.6667){\SetFigFont{5}{6}{rm}.}}
%\put(1683,507){\makebox(1.6667,11.6667){\SetFigFont{5}{6}{rm}.}}
\put(1810,-747){$\ldots$}
\put(1810,-966){$\ldots$}
\put(1810,-1303){$\ldots$}
\put(1810,-1522){$\ldots$}
\put(4388,-1303){$\ldots$}
\put(4388,-966){$\ldots$}
\put(4388,-747){$\ldots$}
\put(4388,-1522){$\ldots$}
\put(4100,-463){$MIN$}
\put(1450,-475){$MAX$}
%\put(5513,-1522){$\mathop {}\nolimits_{\sigma _X }$}
%\put(2965,-1522){$\mathop {}\nolimits_{\sigma _X }$}
\put(2104,-1892){(a)}
\put(4651,-1892){(b)}

\put(5231,-1511){\oval(236, 72)[tr]}
\put(5231,-1511){\oval(236, 72)[tl]}
\put(2678,-1514){\oval(236, 72)[tr]}
\put(2678,-1514){\oval(236, 72)[tl]}

\put(1278,-715){\circle*{90}}%{\oval(84,92)}
\put(1646,-939){\circle*{90}}%{\oval(84,92)}
\put(2206,-1270){\circle*{90}}%{\oval(84,92)}
\put(3002,-1494){\oval(84,92)}
\put(3732,-713){\oval(84,92)}
\put(3831,-716){\circle*{90}}%{\oval(84,92)}
\put(3932,-714){\oval(84,92)}
\put(4101,-932){\oval(84,92)}
\put(4199,-936){\circle*{90}}%{\oval(84,92)}
\put(4300,-931){\oval(84,92)}
\put(4645,-1272){\oval(84,92)}
\put(4749,-1273){\circle*{90}}%{\oval(84,92)}
\put(4863,-1272){\oval(84,92)}
\put(5555,-1494){\oval(84,92)}

\put(5035,-1394){\line( 0,-1){235}}
\put(5035,-1629){\line( 1, 0){376}}
\put(5411,-1629){\line( 0, 1){235}}
\put(5411,-1394){\line(-1, 0){376}}
\put(5120,-1569){\vector( 2, 1){282}}
\put(4100,-977){\line( 0, 1){ 94}}
\put(3001,-1539){\line( 0, 1){ 94}}

\put(5554,-1539){\line( 0, 1){ 94}}
\put(1186,-1610){\framebox(176,215){}}
\put(1554,-1610){\framebox(176,215){}}
\put(1276,-686){\line( 0,-1){705}}
\put(943,-1505){\line( 1, 0){235}}
\put(1364,-1501){\line( 1, 0){188}}
\put(2112,-1614){\framebox(176,215){}}
\put(2288,-1501){\line( 1, 0){188}}
\put(2016,-935){\line( 1, 0){1328}}
\put(2016,-1277){\line( 1, 0){1328}}
\put(1824,-1273){\line(-1, 0){893}}
\put(1824,-935){\line(-1, 0){893}}
\put(1828,-716){\line(-1, 0){893}}
\put(2012,-716){\line( 1, 0){1332}}
\put(2015,-1501){\line( 1, 0){ 94}}
\put(1737,-1499){\line( 1, 0){ 94}}
\put(2202,-1255){\line( 0,-1){141}}
\put(1644,-927){\line( 0,-1){470}}
%\put(1098,-1706){\framebox(2118,1155){}}
\put(2482,-1397){\line( 0,-1){235}}
\put(2482,-1632){\line( 1, 0){376}}
\put(2858,-1632){\line( 0, 1){235}}
\put(2858,-1397){\line(-1, 0){376}}
\put(2567,-1572){\vector( 2, 1){282}}
\put(3739,-1607){\framebox(176,215){}}
\put(4107,-1607){\framebox(176,215){}}
\put(3829,-683){\line( 0,-1){705}}
\put(3917,-1498){\line( 1, 0){188}}
\put(4665,-1611){\framebox(176,215){}}
\put(4841,-1498){\line( 1, 0){188}}
\put(4289,-1498){\line( 1, 0){ 94}}
\put(4569,-1274){\line( 1, 0){1320}}
\put(4377,-1270){\line(-1, 0){893}}
\put(4377,-932){\line(-1, 0){893}}
\put(4381,-713){\line(-1, 0){904}}
\put(4565,-713){\line( 1, 0){1330}}
\put(4749,-1396){\line( 0, 1){141}}
\put(4861,-1321){\line( 0, 1){ 94}}
\put(4298,-978){\line( 0, 1){ 94}}
\put(3931,-763){\line( 0, 1){ 94}}
\put(3732,-760){\line( 0, 1){ 94}}
\put(4197,-922){\line( 0,-1){470}}
\put(4567,-1501){\line( 1, 0){ 94}}
\put(4644,-1315){\line( 0, 1){ 94}}
\put(4569,-932){\line( 1, 0){1326}}
%\put(3642,-1710){\framebox(2118,1155){}}
\put(3740,-1496){\line(-1, 0){254}}
\put(2866,-1496){\line( 1, 0){470}}
\put(5411,-1495){\line( 1, 0){470}}

\end{picture}%

\end{center}
\caption{Quantum circuits for (a)the MAX operator and (b) the MIN operator, followed by a partial measurement then a negation 
to reset the auxiliary qubit $\left| {ax_2 } \right\rangle$. }
\label{mmfig}
\end{figure*}
\end{center}

\subsection{Maximization of the Satisfied Constraints}

Let $\left| {\psi _c } \right\rangle$ be a superposition on $M$ states as follows,

\begin{equation} 
\left| {\psi _c } \right\rangle  = \alpha \sum\limits_{k = 0}^{M - 1} { \left| {z_k } \right\rangle }, 
\end{equation}

\noindent
where each $\left| {z_k } \right\rangle$ is an $m$-qubit state 
and let $d_k=\left\langle {z_k } \right\rangle$ be the number of 1's in state  $\left| {z_k } \right\rangle$ such that 
$\left| {z_k } \right\rangle \ne \left| {0 } \right\rangle^{\otimes m}$, i.e. $d_k \ne 0$. This will be referred to as the 
1-distance of $\left| {z_k } \right\rangle$.

The max-bisection graph $\left| {x_{max}} \right\rangle$ is equivalent to find the state 
$\left| {z_{max}} \right\rangle$ with $d_{max}=max\{d_k,\,0\le k \le M-1\}$ 
and the state $\left| {z_{min} } \right\rangle$ with $d_{min}=min\{d_k,\,0\le k \le M-1\}$ is equivalent to 
the min-bisection graph $\left| {x_{min}} \right\rangle$. 
Finding the state $\left| {z_{min} } \right\rangle$ with the minimum number of 1's is equivalent to 
finding the state with the maximum number of 0's, so, to clear ambiguity, 
let $d_{max1}=d_{max}$ be the maximum number of 1's and $d_{max0}=d_{min}$ be the maximum number of 0's, 
where the number of 0's in $\left| {z_k } \right\rangle$ will be referred to as the 0-distance 
of $\left| {z_k } \right\rangle$. 

To find either $\left| {z_{max} } \right\rangle$ or $\left| {z_{min} } \right\rangle$, 
when $\left| {\psi _c } \right\rangle$ is measured, add an auxiliary qubit $\left| {ax_2 } \right\rangle$ initialized to state 
$\left| 0 \right\rangle$ to the system $\left| {\psi _c } \right\rangle$ as follows,

\begin{equation} 
\begin{array}{l}
\left| {\psi _m } \right\rangle  = \left| {\psi _c } \right\rangle \otimes \left| 0 \right\rangle\\
\,\,\,\,\,\,\,\,\,\,\,\,\,\,= \alpha \sum\limits_{k = 0}^{M - 1} { \left| {z_k } \right\rangle } \otimes \left| 0 \right\rangle.\\
\end{array}
\end{equation}

To main idea to find $\left| {z_{max} } \right\rangle$ is to apply partial negation on 
the state of $\left| {ax_2 } \right\rangle$ entangled with $\left| {z_k } \right\rangle$ based on 
the number of 1's in $\left| {z_k } \right\rangle$, i.e. more 1's in $\left| {z_k } \right\rangle$, 
gives more negation to the state of $\left| {ax_2 } \right\rangle$ entangled with $\left| {z_k } \right\rangle$. 
If the number of 1's in $\left| {z_k } \right\rangle$ is $m$, then 
the entangled state of $\left| {ax_2 } \right\rangle$ will be fully negated. 
The $m^{th}$ partial negation operator is the $m^{th}$ root of $\sigma _X$ and 
can be calculated using diagonalization as follows, 

\begin{equation}
 V=\sqrt[m]{\sigma _X} = \frac{1}{2}\left[ {\begin{array}{*{20}c}
   {1 + t} & {1 - t}  \\
   {1 - t} & {1 + t}  \\
\end{array}} \right],
\end{equation}

\noindent
where $t={\sqrt[m]{{ - 1}}}$, and applying $V$ for $d$ times on a qubit is equivalent to the operator,
 
\begin{equation}
 V^d  = \frac{1}{2}\left[ {\begin{array}{*{20}c}
   {1 + t^d } & {1 - t^d }  \\
   {1 - t^d } & {1 + t^d }  \\
\end{array}} \right],
\end{equation}

\noindent
such that if $d=m$, then $V^m=\sigma _X$. 
To amplify the amplitude of the state $\left| {z_{max} } \right\rangle$, apply the operator MAX 
on $\left| {\psi _m } \right\rangle$ as will be shown later, 
where MAX is an operator on $m+1$ qubits register that applies $V$ conditionally 
for $m$ times on $\left|ax_2 \right\rangle$ 
based on the number of 1's in $\left| {c_0 c_1  \ldots c_{m-1} } \right\rangle$ 
as follows (as shown in Figure \ref{mmfig}(a)),

\begin{equation}
MAX = Cont\_V(c_0 ,ax_2 )Cont\_V(c_1 ,ax_2 ) \ldots Cont\_V(c_{m - 1} ,ax_2),
\end{equation}
\noindent
so, if $d_1$ is the number of $c_l=1$ in $\left| {c_0 c_1  \ldots c_{m-1} } \right\rangle$ then,

\begin{equation}
MAX\left( {\left| {c_0 c_1 ...c_{m - 1} } \right\rangle  \otimes \left| 0 \right\rangle } \right) = \left| {c_0 c_1 ...c_{m - 1} } \right\rangle  \otimes \left( {\frac{{1 + t^{d_1} }}{2}\left| 0 \right\rangle  + \frac{{1 - t^{d_1} }}{2}\left| 1 \right\rangle } \right).
\end{equation}

Amplifying the amplitude of the state $\left| {z_{min} } \right\rangle$ with the minimum number of 
1's is equivalent to amplifying the amplitude of the state with the maximum number of 
0's. To find $\left| {z_{min} } \right\rangle$, apply the operator MIN 
on $\left| {\psi _m } \right\rangle$ as will be shown later, 
where MIN is an operator on $m+1$ qubits register that applies $V$ conditionally 
for $m$ times on $\left|ax_2 \right\rangle$ 
based on the number of 0's in $\left| {c_0 c_1  \ldots c_{m-1} } \right\rangle$ 
as follows (as shown in Figure \ref{mmfig}(b)),

\begin{equation}
MIN = Cont\_V(\overline{c_0} ,ax_2 )Cont\_V(\overline{c_1} ,ax_2 ) \ldots Cont\_V(\overline{c_{m-1}} ,ax_2),
\end{equation}
\noindent
where $\overline{c_l}$ is a temporary negation of $c_l$ before and after the application 
of $Cont\_V(c_l ,ax_2)$ as shown in Figure \ref{mmfig}, 
so, if $d_0$ is the number of $c_l=0$ in $\left| {c_0 c_1  \ldots c_{m-1} } \right\rangle$ then,

\begin{equation}
MIN\left( {\left| {c_0 c_1 ...c_{m - 1} } \right\rangle  \otimes \left| 0 \right\rangle } \right) = \left| {c_0 c_1 ...c_{m - 1} } \right\rangle  \otimes \left( {\frac{{1 + t^{d_0} }}{2}\left| 0 \right\rangle  + \frac{{1 - t^{d_0} }}{2}\left| 1 \right\rangle } \right).
\end{equation}

For the sake of simplicity and to avoid duplication, 
the operator $Q$ will denote either the operator $MAX$ or the operator $MIN$, $d$ will denote either $d_1$ or $d_0$, 
$\left| {z_s } \right\rangle$ will denote either $\left| {z_{max} } \right\rangle$ or $\left| {z_{min} } \right\rangle$, 
and $d_s$ will denote either $d_{max1}$ or $d_{max0}$, so, 

\begin{equation}
Q\left( {\left| {c_0 c_1 ...c_{m - 1} } \right\rangle  \otimes \left| 0 \right\rangle } \right) = \left| {c_0 c_1 ...c_{m - 1} } \right\rangle  \otimes \left( {\frac{{1 + t^{d} }}{2}\left| 0 \right\rangle  + \frac{{1 - t^{d} }}{2}\left| 1 \right\rangle } \right),
\end{equation}

\noindent
and the probabilities of finding the auxiliary qubit $\left|ax_2 \right\rangle$ in state 
${\left| 0 \right\rangle }$ or ${\left| 1 \right\rangle }$ when measured is respectively as follows,

\begin{equation}
\begin{array}{l}
 Pr{(ax_2  = 0)}  = \left| {\frac{{1 + t^d }}{2}} \right|^2  = \cos ^2 \left( {\frac{{d\pi }}{{2m}}} \right), \\ 
 Pr{(ax_2  = 1)}  = \left| {\frac{{1 - t^d }}{2}} \right|^2  = \sin ^2 \left( {\frac{{d\pi }}{{2m}}} \right). \\ 
 \end{array}
\end{equation}

%==============================================

To find the state ${\left| z_s \right\rangle }$ in $\left| {\psi _m } \right\rangle$, 
the proposed algorithm is as follows, as shown in Figure \ref{mmfig}:

\begin{itemize}
\item[1-] Let $\left| {\psi _r } \right\rangle = \left| {\psi _m } \right\rangle$.
\item[2-]  Repeat the following steps for $r$ times,
\begin{itemize}
\item[i-] Apply the operator $Q$ on $\left| {\psi _r } \right\rangle$.
\item[ii-] Measure $\left| ax_2 \right\rangle$, 
if $\left|ax_2 \right\rangle=\left|1 \right\rangle$, 
then let the system post-measurement is $\left| {\psi _r } \right\rangle$, apply $\sigma_X$ on $\left| ax_2 \right\rangle$ 
to reset to $\left| 0 \right\rangle$ for the next iteration and then go to Step (i), otherwise restart the stage and go to Step (1).
\end{itemize}
\item[3-] Measure the first $m$ qubits in $\left| {\psi _r } \right\rangle$ to read $\left| z_s \right\rangle$.
\end{itemize}

For simplicity and without loss of generality, assume that a single $\left| z_s \right\rangle$ exists in $\left| \psi_v \right\rangle$, although such states 
will exist in couples since each $\left| z_s \right\rangle$ is entangled with a variable assignment 
$\left| x_s \right\rangle$ and each $\left| x_s \right\rangle$ is equivalent to 
$\left| \overline {x_s} \right\rangle$, moreover, different variable assignments might give rise to 
constraint vectors with maximum distance, but such information is not known in advance.

Assuming that the algorithm finds $\left|ax_2 \right\rangle=\left|1 \right\rangle$ for $r$ times 
in a row, then the probability of finding $\left|ax_2 \right\rangle=\left|1 \right\rangle$ after Step (2-i) in 
the $1^{st}$ iteration, i.e. $r=1$ is given by,

\begin{equation}
Pr^{(1)}{(ax_2  = 1)}  = \alpha ^2 \sum\limits_{k = 0}^{M - 1} {\sin ^2 \left( {\frac{{d_k \pi }}{{2m}}} \right)}.
\label{probax2}
\end{equation} 

The probability of finding $\left|\psi_r \right\rangle=\left|z_s \right\rangle$ after Step (2-i) in 
the $1^{st}$ iteration, i.e. $r=1$ is given by,

\begin{equation}
Pr^{(1)}{(\psi_{r}  = z_s)}  = \alpha ^2 {\sin ^2 \left( {\frac{{d_s \pi }}{{2m}}} \right)} .
\end{equation}

The probability of finding $\left|ax_2 \right\rangle=\left|1 \right\rangle$ after Step (2-i) in the $r^{th}$ iteration, i.e. $r>1$ is given by,

\begin{equation}
Pr^{(r)}{(ax_2  = 1)}  = \frac{{\sum\limits_{k = 0}^{M - 1} {\sin ^{2r} \left( {\frac{{d_k \pi }}{{2m}}} \right)} }}{{\sum\limits_{k = 0}^{M - 1} {\sin ^{2(r - 1)} \left( {\frac{{d_k \pi }}{{2m}}} \right)} }}.
\end{equation}

%\begin{figure}[htbp]
%\vspace*{7cm}
%\centerline{\includegraphics{fig1.eps}}
%\caption{ The probability of success.. ini=0.5411, fin=0.7263, n=20;e=97} 
%\label{fig1}
%\end{figure}

%\begin{figure}[htbp]
%\vspace*{7cm}
%\centerline{\includegraphics{fig2.eps}}
%\caption{ The probability of success.. ini=0.9571, fin=0.9765, n=20;e=97} 
%\label{fig1}
%\end{figure}

\begin{figure}[htbp]
%\vspace*{7cm}
\centerline{\includegraphics{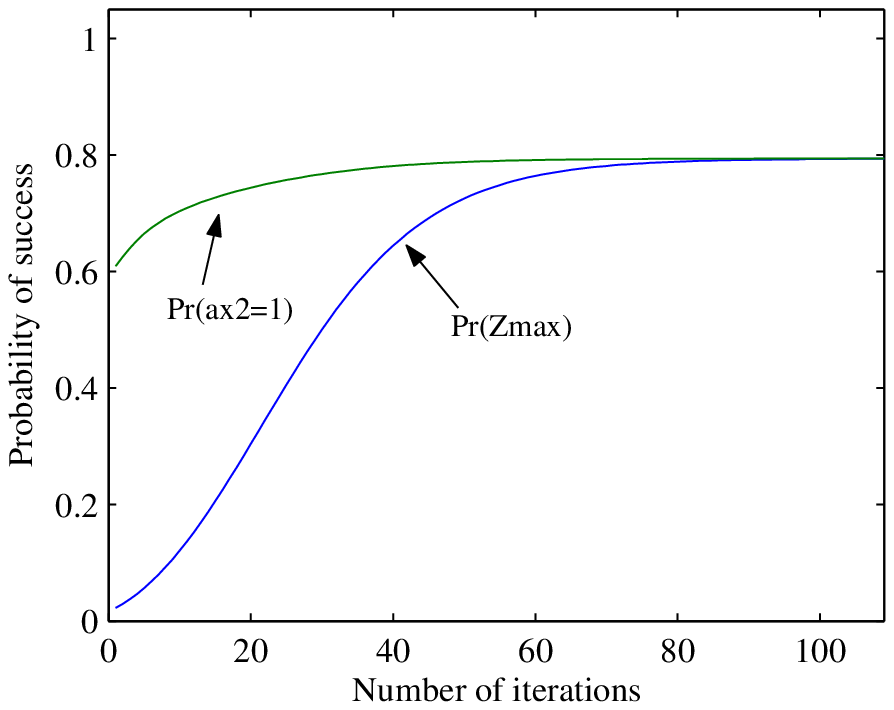}}
\caption{The probability of success for a max-bisection instance of the graph shown in Figure \ref{graphex} with $n=8$ and $m=12$ where 
 the probability of success of $\left|ax_2\right\rangle$ is 0.6091 after the first iteration and  
with probability of success of 0.7939 after iterating the algorithm where the probability of success of $\left|z_{max}\right\rangle$ 
is amplified to reach the probability of success of $\left|ax_2\right\rangle$.} 
\label{fig21}
\end{figure}

The probability of finding $\left|\psi_r \right\rangle=\left|z_s \right\rangle$ after Step (2-i) in the $r^{th}$ iteration, i.e. $r>1$ is given by,

\begin{equation}
Pr^{(r)}{(\psi_{r}  = z_s)}  = \frac{{{\sin ^{2r} \left( {\frac{{d_s \pi }}{{2m}}} \right)} }}{{\sum\limits_{k = 0}^{M - 1} {\sin ^{2(r - 1)} \left( {\frac{{d_k \pi }}{{2m}}} \right)} }}.
\end{equation} 

To get the highest probability of success for $Pr{(\psi_{r}  = z_s)}$, Step (2) should be repeated until, 
$\left| Pr^{(r)}{(ax_2  = 1)}  - Pr^{(r)}{(\psi_r  = z_s)} \right| \le \epsilon$ for small $\epsilon \ge 0$ as shown in Figure \ref{fig21}. 
This happens when 
$\sum\nolimits_{k = 0,k\ne s}^{M - 1} {\sin ^{2r} \left( {{\textstyle{{d_k \pi } \over {2m}}}} \right)}  \le \epsilon$. 
Since the Sine function is a decreasing function then for sufficient large $r$,

\begin{equation}
\sum\limits_{k = 0,k\ne s}^{M - 1} {\sin ^{2r} \left( {\frac{{d_k \pi }}{{2m}}} \right)}  \approx \sin ^{2r} \left( {\frac{{d_{ns} \pi }}{{2m}}} \right),
\end{equation} 
\noindent
where $d_{ns}$ is the next maximum distance less than $d_s$. The values of $d_s$ and $d_{ns}$ are unknown in advance, so let $d_s=m$ be the number of edges, 
then in the worst case when $d_s=m$, $d_{ns}=m-1$ and $m=n(n-1)/2$, the required number of iterations $r$ 
for $\epsilon  = 10^{ - \lambda }$ and $\lambda>0$ can be calculated using the formula,

\begin{equation}
0 < \sin ^{2r} \left( {\frac{{(m-1) \pi }}{{2m}}} \right) \le \epsilon,
\end{equation} 

\begin{equation}
\begin{array}{l}
 r \ge \frac{{\log \left( \epsilon  \right)}}{{2\log \left( {\sin \left( {\frac{{\left( {m - 1} \right)\pi }}{{2m}}} \right)} \right)}} \\ 
 \,\,\,\, = \frac{{\log \left( {10^{ - \lambda } } \right)}}{{2\log \left( {\cos \left( {\frac{\pi }{{2m}}} \right)} \right)}} \\ 
 \,\,\,\, \ge \lambda \left( {\frac{{2m}}{\pi }} \right)^2  \\
 \,\,\,\,=O\left( {m^2 } \right),\\
 \end{array}
\end{equation} 
\noindent
where $0 \le m \le {\textstyle{{n(n - 1)} \over 2}}$. For a complete graph where $m={\textstyle{{n(n - 1)} \over 2}}$, then the upper bound 
for the required number of iterations $r$ is $O\left( {n^4 } \right)$. Assuming that 
a single $\left|z_s \right\rangle$ exists in the superposition will increase the required number of iterations, 
so it is important to notice here that the probability of success will not be over-cooked by increasing the 
required number of iteration $r$ similar to the common amplitude 
amplification techniques.

\subsection{Adjustments on the Proposed Algorithm}

During the above discussion, two problems will arise during the implementation of the proposed algorithm.
The first one is to finding $\left|ax_2 \right\rangle=\left|1 \right\rangle$ for $r$ times in a row 
which a critical issue in the 
success of the proposed algorithm to terminate in polynomial time. The second problem is that the value of $d_s$ is not known in 
advance, where the value of $Pr^{(1)}{(ax_2  = 1)}$ 
shown in Eqn. \ref{probax2} plays an important role in the success of finding 
$\left|ax_2 \right\rangle=\left|1 \right\rangle$ in the next iterations, this value depends heavily on 
the density of 1's, i.e. the ratio ${\textstyle{{d_s } \over m}}$.

Consider the case of a complete graph with even number of vertices, 
where the number of egdes $m = {\textstyle{{n(n - 1)} \over 2}}$ and all $\left|z_k \right\rangle$'s are equivalent 
and each can be taken as $\left|z_s \right\rangle$ then, 

\begin{equation}
Pr^{(1)}{(ax_2  = 1)}  =  M \alpha ^2 {\sin ^2 \left( {\frac{{d_s\pi }}{{2m}}} \right)}.
\label{probax2_2}
\end{equation}

This case is an easy case where setting $m=d_s$ in $m^{th}$ root of $\sigma _X$ will lead to a probability of 
success of certainty after a single iteration. Assuming a blind approach where $d_{s}$ is not known, 
then this case represents the worst ratio ${\textstyle{{d_s } \over m}}$ where the probability of success will be $\approx0.5$ for 
sufficient large graph. Iterating the algorithm will not lead to any increase in the probability of both 
$\left|z_s \right\rangle$ and $\left|ax_2 \right\rangle$.

In the following, adjustments on the proposed algorithm for the max-bisection and the min-bisection graph will be presented 
to overcome these problems, i.e. to be able to find $\left|ax_2 \right\rangle=\left|1 \right\rangle$ 
after the first iteration with the highest probability of success 
without a priori knowledge of $d_s$.

\begin{figure}[htbp]
%\vspace*{7cm}
\centerline{\includegraphics{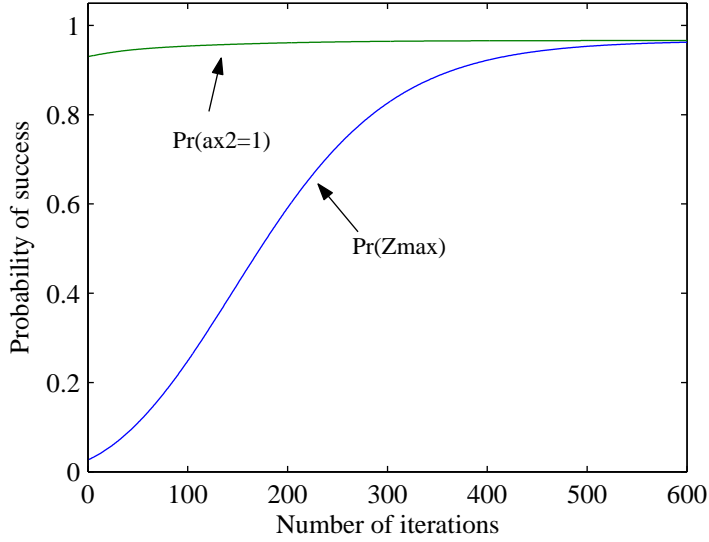}}

\caption{The probability of success for a max-bisection instance of the graph shown in Figure \ref{graphex} with $n=8$, 
$m=12$, $\mu_{max}=31$ and $\delta=0.9$, where the probability of success of $\left|ax_2\right\rangle$ is 0.9305 
after the first iteration and with probability of success of 0.9662 after iterating the algorithm where the probability of success 
of $\left|z_{max}\right\rangle$ is amplified to reach the probability of success of $\left|ax_2\right\rangle$.} 
\label{fig22}
\end{figure}

\subsubsection*{Adjustment for the Max-Bisection Problem}

In an arbitrary graph, the density of 1's will be ${\textstyle{{d_{max1} } \over m}}$.
In the case of a complete graph, there are $M$ states with 1-distance ($d_k$) equals to ${\textstyle{{n^2} \over 4}}$. 
This case represents the worst density of 1's where the density will be 
${\textstyle{{n^2 } \over {2n(n - 1)}}}$  slightly greater than 0.5 for arbitrary large $n$. 
Iterating the proposed algorithm will not amplify the amplitudes after arbitrary number of iterations.  
To overcome this problem, add 
$\mu_{max}$ temporary qubits initialized to state $\left|1 \right\rangle$ 
to the register $\left| {c_0 c_1 ...c_{m - 1} } \right\rangle$ as follows,

\begin{equation}
\left| {c_0 c_1 ...c_{m - 1} } \right\rangle \to \left| {c_0 c_1 \ldots c_{m - 1}c_{m}c_{m+1}\ldots c_{m+\mu_{max}-1} } \right\rangle,
\end{equation}
\noindent
so that the extended number of edges $m_{ext}$ will be $m_{ext}=m+\mu_{max}$ and $V=\sqrt[{m_{ext} }]{{\sigma _X }}$ will be used instead of 
$V=\sqrt[{m}]{{\sigma _X }}$ in the MAX operator, then the density of 1's will be 
${\textstyle{{n^2  + 4\mu _{max } } \over {2n(n - 1) + 4\mu _{max } }}}$. To get a probability of success $Pr_{max}$ 
to find $\left|ax_2 \right\rangle=\left|1 \right\rangle$ after the first iteration,

\begin{equation}
Pr ^{(1)} {\left( {ax_2  = 1} \right)} = M \alpha ^2 \sin ^2 \left( {\frac{{\pi \left( {{\textstyle{{n^2 } \over 4}} + \mu _{\max } } \right)}}{{2\left( {{\textstyle{{n(n - 1)} \over 2}} + \mu _{\max } } \right)}}} \right) \ge Pr_{\max }, 
\end{equation}
\noindent
then the required number of temporary qubits $\mu_{max}$ is calculated as follows,

\begin{equation}
\mu _{\max }  \ge \frac{1}{{1 - \omega }}\left( {\frac{{n^2 }}{2}\left( {2\omega  - 1} \right) - \frac{n}{2}\omega } \right),
\end{equation}
\noindent
where $\omega  = {\textstyle{2 \over \pi }}\sin ^{ - 1} \left( {\sqrt {{{{Pr_{\max } } \over {M\alpha ^2 }}}} } \right)$ 
and $Pr_{\max }  < {\textstyle{{M \alpha ^2 }}}$, with $M\alpha^2=1$ so let 
$Pr_{\max }  = \delta {\textstyle{{M \alpha ^2 } }}$ such that $0<\delta<1$. For example, if $\delta=0.9$, 
then $Pr ^{(1)} \left( {ax_2  = 1} \right)$ will be at least 90$\%$ as shown in Figure \ref{fig22}. To conclude, the problem of low 
density of 1's can be solved with a polynomial increase in the number of qubits to get the solution 
$\left|z_{max} \right\rangle$ in $O\left(m_{ext}^2\right)=O\left( {n^4 } \right)$ iterations with arbitrary 
high probability $\delta<1$ to terminate in poly-time, i.e. to read $\left|ax_2 \right\rangle=\left|1 \right\rangle$ 
for $r$ times in a row.

\subsubsection*{Adjustment for the Min-Bisection Problem}

Similar to the above approach, in an arbitrary graph, the density of 0's will be ${\textstyle{{d_{max0} } \over m}}$. 
In the case of a complete graph, there are $M$ states with 0-distance ($d_k$) equals to 
${\textstyle{{n(n - 1)} \over 2}}-{\textstyle{{n^2} \over 4}}$. 
This case represents the worst density of 0's where the density will be ${\textstyle{{n-2 } \over {2(n - 1)}}}$ 
slightly less than 0.5 for arbitrary large $n$. Iterating the proposed algorithm will not lead to any amplification 
after arbitrary number of iterations. To overcome this problem, add $\mu_{min}$ temporary qubits initialized to 
state $\left|0 \right\rangle$ to the register $\left| {c_0 c_1 ...c_{m - 1} } \right\rangle$ as follows,

\begin{equation}
\left| {c_0 c_1 ...c_{m - 1} } \right\rangle \to \left| {c_0 c_1 \ldots c_{m - 1}c_{m}c_{m+1}\ldots c_{m+\mu_{min}-1} } \right\rangle,
\end{equation}
\noindent
so that the extended number of edges $m_{ext}$ will be $m_{ext}=m+\mu_{min}$ and $V=\sqrt[{m_{ext} }]{{\sigma _X }}$ 
will be used instead of $V=\sqrt[{m}]{{\sigma _X }}$ in the MIN operator, then the density of 0's will be 
${\textstyle{{n^2 -2n  + 4\mu _{min } } \over {2n(n - 1) + 4\mu _{min } }}}$. To get a probability of success $Pr_{max}$ 
to find $\left|ax_2 \right\rangle=\left|1 \right\rangle$ after the first iteration,

\begin{equation}
Pr ^{(1)} \left( {ax_2  = 1} \right) = M \alpha ^2 \sin ^2 \left( {\frac{{\pi \left( {{\textstyle{{n(n - 1)} \over 2}} - {\textstyle{{n^2 } \over 4}} + \mu _{min } } \right)}}{{2\left( {{\textstyle{{n(n - 1)} \over 2}} + \mu _{min } } \right)}}} \right) \ge Pr_{\max },
\end{equation}
\noindent
then the required number of temporary qubits $\mu_{min}$ is calculated as follows,

\begin{equation}
\mu _{\min }  \ge \frac{{n^2 }}{4}\left( {\frac{{2\omega  - 1}}{{1 - \omega }}} \right) + \frac{n}{2},
\end{equation}
\noindent
where $\omega  = {\textstyle{2 \over \pi }}\sin ^{ - 1} \left( {\sqrt {{{{Pr_{\max } } \over {M\alpha ^2 }}}} } \right)$ 
and $Pr_{\max }  < {\textstyle{{M \alpha ^2 }}}$, with $M\alpha^2=1$ so let 
$Pr_{\max }  = \delta {\textstyle{{M \alpha ^2 } }}$ such that $0<\delta<1$. For example, if $\delta=0.9$, 
then $Pr ^{(1)} \left( {ax_2  = 1} \right)$ will be at least 90$\%$. 
To conclude similar to the case of the max-bisection graph, the problem of low density of 0's can be solved with a polynomial 
increase in the number of qubits, larger than the case of the max-bisection graph, to get the solution $\left|z_{min} \right\rangle$ 
in $O\left(m_{ext}^2\right)=O\left( {n^4 } \right)$ iterations with arbitrary high probability $\delta<1$ 
to terminate in poly-time, i.e. to read $\left|ax_2 \right\rangle=\left|1 \right\rangle$ 
for $r$ times in a row.

%=============================

\section{Conclusion}

Given an undirected graph $G$ with even number of vertices $n$ and $m$ unweighted edges. 
The paper proposed a BQP algorithm to solve the max-bisection problem and the min-bisection problem, where a general graph 
is considered for both problems.
 
The proposed algorithm uses a representation of the two problems as a Boolean constraint satisfaction problem,  
where the set of edges of a graph are represented as a set of constraints. The algorithm is divided into three 
stages, the first stage prepares a superposition of all possible equally sized graph partitions in $O\left( {\sqrt[4]{n}} \right)$ 
using an amplitude amplification technique that runs in $O\left( {\sqrt {{\textstyle{N \over M}}} } \right)$, 
for $N=2^n$ and $M$ is the number of possible graph partitions. The algorithm, in the second stage, 
evaluates the set of constraints for all possible graph partitions. In the third stage, the algorithm amplifies 
the amplitudes of the best graph bisection that achieves maximum/minimum satisfaction to the set of constraints 
using an amplitude amplification technique that applies an iterative partial negation 
where more negation is given to the set of constrains with more satisfied constrains and a partial measurement to amplify 
the set of constraints with more negation. 
The third stage runs in $O(m^2)$ and in the worst case runs in $O(n^4)$ for a dense graph. 
It is shown that the proposed algorithm achieves an arbitrary high probability of success of 
$1-\epsilon$ for small $\epsilon>0$ using a polynomial increase in the space resources by adding dummy constraints with predefined 
values to give more negation to the best graph bisection.


\begin{thebibliography}{10}

\bibitem{Ref1} 
Armbruster, M.,
Branch-and-cut for a Semidefinite Relaxation of Large-scale Minimum Bisection Problems. 
Ph.D. thesis, Technische Universität Chemnitz, 2007.

\bibitem{Ref3} 
Armbruster, M., Fügenschuh, M., Helmberg, C. and Martin, A.,
A comparative Study of Linear and Semidefinite Branch-and-cut Methods for Solving the Minimum Graph Bisection Problem. 
In: Proceedings ofthe Conference on Integer Programming and Combinatorial Optimization (IPCO), LNCS, vol. 5035, pp. 112-–124, 2008.


\bibitem{BAZ05} 
Bazgan, C. and Karpinski, M., 
On the Complexity of Global Constraint Satisfaction.
Proceeding of 16$^{th}$ Annual International Symposium on Algorithms and Computation,
vol. LNCS 3827, Springer-Verlag, pp. 624–-633, 2005.

\bibitem{CombOptBook}
Bazgan, C., 
Paradigms of Combinatorial Optimization: Problems and New Approaches. 
(Chapter 1: Optimal Satisfiability), vol. 2, p.25,    
Vangelis Th. Paschos (Editor),  Wiley-ISTE, 2010.

\bibitem{Ref8} 
Bhatt, S.N. and Leighton, F.T.,
A Framework for Solving VLSI Graph Layout Problems. 
Journal of Computer and System Sciences, 28(2), pp. 300-–343, 1984.

\bibitem{Ref16} 
Delling, D., Goldberg, A.V., Razenshteyn, I. and Werneck, R.F.,
Graph Partitioning with Natural Cuts. 
In: Proceedings of the IEEE International Parallel and Distributed Processing Symposium (IPDPS), pp. 1135–-1146. IEEE, 2011.

\bibitem{Ref19} 
Delling, D. and Werneck, R.F.,
Faster Customization of Road Networks. 
In: Proceedings of the International Symposium on Experimental Algorithms (SEA), LNCS, vol. 7933, 
pp. 30–-42. Springer, Berlin, Heidelberg, 2013.

\bibitem{Delling2014} 
Delling, D., Fleischman, D., Goldberg, A. V., Razenshteyn, I. and Werneck R. F.,
An Exact Combinatorial Algorithm for Minimum Graph Bisection. 
Mathematical Programming, DOI 10.1007/s10107-014-0811-z, 2014.

\bibitem{NPhard2007} 
Josep Díaza, J. and Kamińskib, M., 
MAX-CUT and MAX-BISECTION are NP-hard on unit disk graphs
Theoretical Computer Science, (377):1–3,pp. 271-–276, 2007.

\bibitem{Feige2006}
Feige, U. and Langberg, M., 
The RPR$^2$ Rounding Technique for Semidefinite Programs. 
Journal of Algorithms, 60: pp. 1–-23, 2006.

\bibitem{Ref22} 
Feldmann, A.E. and Widmayer, P.,
An $O(n^4)$ Time Algorithm to Compute the Bisection Width of Solid Grid Graphs. 
In: Proceedings of the European Symposium on Algorithms (ESA), LNCS, vol. 6942, pp. 143–-154, Springer, 2011.





\bibitem{Guruswami2011}
Guruswami, V., Makarychev, Y., Raghavendra, P., Steurer, D. and Zhou, Y.,
Finding Almost-Perfect Graph Bisections. 
In: Proceedings of ICS, pp. 321–-337, 2011.


\bibitem{Ref25} 
Garey, M.R. and Johnson, D.S., 
Computers and Intractability. A Guide to the Theory of NP-Completeness. W. H. Freeman and Company, London, 1979.

\bibitem{Ref26} 
Garey, M.R., Johnson, D.S. and Stockmeyer, L.J., 
Some Simplified NP-Complete Graph Problems. 
Theoretical Computer Science, 1, pp. 237–-267, 1976.

\bibitem{Grover1997}
Grover, L. K.,
Quantum Mechanics Helps in Searching for a Needle in a Haystack.
Physical Review Letters, 79, 325, 1997.


\bibitem{Ref30} 
Hager, W.W., Phan, D.T. and Zhang, H.,
An Exact Algorithm for Graph Partitioning. 
Mathematical Programming, 137, pp. 531–-556, 2013.

\bibitem{Ref33} 
Hendrickson, B. and Leland, R.,
An Improved Spectral Graph Partitioning Algorithm for mapping Parallel Computations. 
SIAM Journal on Scientific Computing, 16(2), pp.452–-469, 1995.


\bibitem{Ref31} 
Hein, M. and Bühler, T., 
An Inverse Power Method for Nonlinear Eigen Problems with Applications in 1-spectral Clustering and Sparse PCA. 
In: Proceedings Advances in Neural Information Processing Systems (NIPS), pp. 847–-855, 2010.

\bibitem{hoyer00} 
H{\o}yer, P., 
Arbitrary Phases in Quantum Amplitude Amplification,
Physical Review A, 62, 52304, 2000.


\bibitem{Ref37} 
Jansen, K., Karpinski, M., Lingas, A. and Seidel, E.,
Polynomial Time Approximation Schemes for MAX-BISECTION on Planar and Geometric Graphs. 
SIAM Journal on Computing, 35, pp. 110-–119, 2005.

\bibitem{Ref39} 
Johnson, E., Mehrotra, A., Nemhauser, G., 
Min-Cut Clustering. 
Mathematical Programming, 62, pp. 133-–152, 1993.

\bibitem{Ref46} 
Kwatra, V., Schödl, A., Essa, I., Turk, G. and Bobick, A.,
Graph Cut Textures: Image and Video Synthesis Using Graph Cuts. 
ACM Transactions on Graphics, 22, pp.277-–286, 2003.

\bibitem{Ref47} 
Land, A.H. and Doig, A.G.,
An Automatic Method of Solving Discrete Programming Problems. 
Econometrica 28(3), pp. 497–-520, 1960.

\bibitem{Ref50} 
Lipton, R.J. and Tarjan, R., 
Applications of a Planar Separator Theorem. 
SIAM Journal on Computing, 9, pp.615–-627, 1980.

\bibitem{Ref51} 
Malewicz, G., Austern, M.H., Bik, A.J., Dehnert, J.C., Horn, I., Leiser, N. and Czajkowski, G., 
Pregel, A System for Large-Scale Graph Processing. 
In: PODC, pp. 6. ACM, 2009.

\bibitem{Ref53} 
Meyerhenke, H., Monien,B. and Sauerwald, T.,
A New Diffusion-based Multilevel Algorithm for Computing Graph Partitions. 
Journal of Parallel and Distributed Computing, 69(9), pp. 750-–761, 2009.


\bibitem{Ref55} 
Räcke, H.,
Optimal Hierarchical Decompositions for Congestion Minimization in Networks. 
In: Proceedings of the ACM Symposium on Theory of Computing (STOC), pp. 255-–263. ACM Press, New York, 2008. 

\bibitem{Raghavendra2012}
Raghavendra, P. and Tan, N.,
Approximating CSPs with Global Cardinality Constraints Using SDP Hierarchies. 
In: Proceedings of SODA, pp. 373–-387, 2012.



\bibitem{ChebPoly}
  Rivlin, T. J.,
  Chebyshev Polynomials.
  Wiley, New York, 1990.

\bibitem{Ref62} 
Shi, J. and Malik, J.,
Normalized Cuts and Image Segmentation. 
IEEE Transactions on Pattern Analysis and Machine Intelligence, 22(8), pp.888–-905, 2000.
 




\bibitem{Ref58} 
Sanders, P. and Schulz, C.,
Distributed Evolutionary Graph Partitioning. 
In: Proceedings of the Algorithm Engineering and Experiments (ALENEX), pp. 16-–29. SIAM, 2012.

\bibitem{Ref59} 
Sanders, P. and Schulz, C.,
Think Locally, Act Globally: Highly Balanced Graph Partitioning. 
In: Proceedings of the International Symposium on Experimental Algorithms (SEA), LNCS, vol. 7933, pp. 164–-175. Springer, Berlin, Heidelberg, 2013.


\bibitem{ZiXu2014} 
Xu, Z., Du, D. and Xu D., 
Improved Approximation Algorithms for the Max-Bisection and the Disjoint 2-Catalog Segmentation Problems. 
Journal of Combinatorial Optimization, (27):2, pp. 315--327, 2014.


\bibitem{Younes2007}
Younes, A., Rowe, J. and Miller, J.,
Enhanced Quantum Searching via Entanglement and Partial Diffusion. 
Physica D. Vol. 237(8) pp. 1074--1078, 2007.

\bibitem{Younes2013}
Younes, A., 
Towards More Reliable Fixed Phase Quantum Search Algorithm.
Applied Mathematics $\&$ Information Sciences.Vol 7, No. 1, pp. 93--98, 2013.




\end{thebibliography}
\end{document}